\documentclass[%
reprint,
superscriptaddress,
amsmath,amssymb,
aps,
prb, 
]{revtex4-2}

\usepackage{booktabs}
\usepackage{verbatim}
\usepackage{textcomp,gensymb}
\usepackage{graphicx}
\usepackage{dcolumn}
\usepackage{bm}
\usepackage{hyperref}
\usepackage[mathlines]{lineno}
\usepackage[normalem]{ulem} 

\usepackage{tabularx}
\usepackage{multirow}
\newcolumntype{Y}{>{\centering\arraybackslash}X}
\newcolumntype{s}{>{\hsize=.8\hsize\centering\arraybackslash}X}  
\newcolumntype{B}{>{\hsize=1.5\hsize}c}  
\newcommand{\fullprof}{\texttt{FullProf Suite}}
\newcommand{\maxmagn}{\texttt{MAXMAGN}}
\newcommand{\musrfit}{\texttt{MUSRFIT}}
\newcommand{\jana}{\texttt{JANA2020}}
\newcommand{\pycrystalfield}{\texttt{PyCrystalField}}
\newcommand{\spinw}{\texttt{SPINW}}
\newcommand{\ce}{Ce$_5$CoGe$_2$}
\newcolumntype{l}{>{\hsize=.8\hsize\raggedright\arraybackslash}X}
\newcolumntype{L}{>{\hsize=\hsize\raggedright\arraybackslash}X}

\usepackage{color}

\begin{document}
	
	\title{Non-collinear ferromagnetism in the Kondo lattice Ce$_5$CoGe$_2$}
	
	\author{Jinyu Wu}
	\affiliation{Center for Correlated Matter and School of Physics, Zhejiang University, Hangzhou 310058, China.}
	\author{Jiawen Zhang}
	\affiliation{Center for Correlated Matter and School of Physics, Zhejiang University, Hangzhou 310058, China.}
	\author{Toni Shiroka}
	\affiliation{Laboratory for Solid State Physics, ETH Zürich, 8093 Zürich, Switzerland.}
	\affiliation{PSI Center for Neutron and Muon Sciences CNM, 5232 Villigen PSI, Switzerland.}
	\author{Shams Sohel Islam}
	\affiliation{PSI Center for Neutron and Muon Sciences CNM, 5232 Villigen PSI, Switzerland.}
	\author{Mingyi Wang}
	\affiliation{Center for Correlated Matter and School of Physics, Zhejiang University, Hangzhou 310058, China.}
	\author{Yongjun Zhang}
	\affiliation{Institute for Advanced Materials, Hubei Normal University, Huangshi 435002, China.}
	\author{Devashibhai T. Adroja}
	\affiliation{ISIS Facility, STFC, Rutherford Appleton Laboratory, Chilton, Didcot, Oxfordshire OX11 0QX, United Kingdom}
	\affiliation{Highly Correlated Matter Research Group, Physics Department, University of Johannesburg, PO Box 524, Auckland Park 2006, South Africa.}
	\author{Yu Liu}
	\affiliation{Center for Correlated Matter and School of Physics, Zhejiang University, Hangzhou 310058, China.}
	\author{Huiqiu Yuan}
	\affiliation{Center for Correlated Matter and School of Physics, Zhejiang University, Hangzhou 310058, China.}
	\affiliation{Institute for Advanced Study in Physics, Zhejiang University, Hangzhou 310058, China.}
	\affiliation{Institute of Fundamental and Transdisciplinary Research, Zhejiang University, Hangzhou 310058, China}
	\affiliation{State Key Laboratory of Silicon and Advanced Semiconductor Materials, Zhejiang University, Hangzhou 310058, China.}
	\author{Michael Smidman}
	\email{msmidman@zju.edu.cn}
	\affiliation{Center for Correlated Matter and School of Physics, Zhejiang University, Hangzhou 310058, China.}

	\date{\today}
	
\begin{abstract}
The dense Kondo lattice \ce\ exhibits  superconductivity once the magnetic ordering is suppressed by pressure. Here the ambient pressure magnetic state is investigated via magnetization, heat capacity, powder neutron diffraction, and muon spin relaxation ($\mu$SR) measurements. Neutron diffraction results reveal a noncollinear ferromagnetic  structure, where the four inequivalent Ce sites exhibit different magnetic moments.  Point-charge model calculations of the  crystalline-electric field (CEF) ground states corroborate different moments between the sites, and suggest sizeable components of the moments along different directions, consistent with the non-collinear structure. Analysis of the Dzyaloshinskii–Moriya (DM) interaction for the bonds connecting Ce atoms demonstrates that most of these bonds exhibit a nonzero DM vector, suggesting that competition between intersite magnetic exchange interactions, CEF driven single-ion anisotropy, the Kondo effect and the DM interaction may drive the non-collinear ferromagnetism.
\end{abstract}	
	
\maketitle

\section{Introduction}\label{intro}

Cerium-based intermetallic compounds exhibit competition between the Ruderman-Kittel-Kasuya-Yosida (RKKY)  interaction which promotes magnetic order, and the Kondo effect which favors the formation of non-magnetic Kondo singlets. As a result, non-thermal control parameters such as pressure, magnetic-fields or doping can often suppress an antiferromagnetic (AFM) transition to a zero-temperature  quantum critical point (QCP), which are often accompanied by various emergent phenomena such as non-Fermi liquids and unconventional superconductivity \cite{QCP1,QCP2,Weng2016}.

In contrast, in ferromagnets QCPs are not commonly observed, where CeRh$_6$Ge$_4$ stands out as a rare example where a ferromagnetic (FM) QCP can be induced by both applying hydrostatic pressure and doping \cite{shenbin164,zhan2025}, which is accompanied by strange metal behavior in the resistivity and specific heat. In most cases, continuous FM QCPs are avoided \cite{RMPFMQCP}, where instead the FM transition becomes first-order, as in UGe$_2$, URhGe and UCoGe \cite{firstorderQCP}, or there is a change of ground state to AFM order \cite{CeRuPO,Friedemann2018,Sidorov2003} or a Kondo cluster spin-glass \cite{SG1,SG2,SG3}. As such there is particular interest in determining the key factors governing the outcomes of ferromagnetic quantum phase transitions, whereby the role of spin-orbit coupling \cite{Ce164Kirk,Ce164Miser}, nature of the Kondo hybridization and quantum criticality \cite{Ce164Pei,Ce164Piers,Ce164Wang,Ce164WuYi,shenbin164,zhan2025}, electronic structure and crystalline-electric field (CEF) effects \cite{Ce164Wangan,Ce164Shujianwei,Ce164Thomas,Yamamoto2025,Itokazu2025}, and influence of disorder \cite{Belitz1999,Ce164Xu,Ce164Yongjun} have been discussed in relation to allowing for the realization of FM QCPs.

 In the case of a change from FM to AFM ground state, this has been theoretically studied in the context of itinerant ferromagnetism \cite{Chubukov2004,Conduit2009,Karahasanovic2012}, and the interplay of itinerant FM and spin-density wave orders has been proposed to induce quantum tricritical points \cite{Friedemann2018}, motivating detailed characterizations of the ordered state of quantum ferromagnets. Along these lines, the ferromagnet CeAgSb$_2$ exhibits a pressure-induced change to probable AFM order \cite{Sidorov2003}, but while there have been suggestions of AFM components in the ambient pressure magnetic state \cite{Jobiliong2005,Muro1997}, ambient pressure neutron scattering results are well accounted for by uniform collinear FM order together with FM magnetic exchange interactions \cite{andre2000,Araki2003,Nikitin2021}. Meanwhile the itinerant FM state of  LaCrGe$_3$ has been reported to transform to either AFM order \cite{Taufour2016,Kaluarachchi2017} or short-range ordered FM clusters under pressure \cite{Gati2021}, and neutron diffraction suggests a  transition to a canted FM structure at low temperatures \cite{cadogan2013}.

Crystallizing in an orthorhombic structure (space group $Pnma$), Ce$_5M$Ge$_2$  ($M$=Co, Ru, Rh, Ir, Pd) are dense Kondo lattice compounds exhibiting  complex magnetic properties \cite{Ce5121st,Ce5RhGe2,Ce5RuGe2,Ce5PdGe2,Ce5CoGe2poly,Ce5CoGe2single}.  Within the orthorhombic unit cell, the Ce atoms occupy four crystallographically inequivalent sites,  three $4c$ sites and one $8d$ site, and  ab-initio calculations reveal that these four distinct Ce sites likely possess different magnetic moments \cite{Ce5RhGe2,Ce5RuGe2,Ce5PdGe2}.  Ce$_5$RhGe$_2$ is reported to undergo a transition to weak FM order below around 11.3 K \cite{Ce5RhGe2}, while both Ce$_5$RuGe$_2$ \cite{Ce5RuGe2} and Ce$_5$PdGe$_2$ \cite{Ce5PdGe2} transition to ferrimagnetic ordering, and exhibit metamagnetic transitions to different field-induced phases in applied magnetic fields. Meanwhile all three materials have  spin-glass behavior that coexists with the magnetic order below  around 3 to 5 K, well below the magnetic ordering temperature. Recently, single crystals of \ce\  have been synthesized, which exhibits a FM transition at $T_{\textrm{C}}$ = 10.9 K, at which there is also the onset of spin-glass behavior, as evidenced by a frequency-dependent cusp in the ac susceptibility \cite{Ce5CoGe2single}, and there is no sign of the metamagnetic transitions observed in Ce$_5$RuGe$_2$ and Ce$_5$PdGe$_2$ \cite{Ce5RuGe2,Ce5PdGe2}. The application of pressure first leads to a change of ground state from ferromagnetic to antiferromagnetic, which  is suppressed to a QCP upon further increasing the pressure \cite{zhang2026pressureinducedsuperconductivitymagneticquantum}. Meanwhile superconductivity emerges at higher pressures, where the separation of the superconductivity from the magnetic instability in the phase diagram suggests a pairing mechanism beyond one purely mediated by spin fluctuations. Therefore, investigating the magnetic ground state of \ce\ at ambient pressure is important for  both gaining a microscopic understanding of the nature of the magnetic ordering in Kondo ferromagnets and for elucidating the intricate pressure evolution of the different ordered phases in \ce.

In this paper, we investigate the magnetic ground state of \ce\ using neutron diffraction and muon-spin relaxation ($\mu$SR). Neutron diffraction measurements reveal a propagation vector $\textit{\textbf{k}}=0$, which is consistent with the FM behavior reported previously \cite{Ce5CoGe2poly,Ce5CoGe2single}. Analysis of the diffraction results reveal that the Ce-magnetic moments form a non-collinear FM structure, with different magnetic moments on the crystallographically inequivalent Ce-sites, which is corroborated by point-charge model calculations of the CEF ground state. $\mu$SR measurements confirm the presence of long-range magnetic order, where coherent oscillations appear in the spectra below $T_{\textrm{C}}$, and the temperature dependence of the internal fields is consistent with mean-field behavior.

\section{Experimental details\label{exp detail}}

Single crystals of Ce$_5$CoGe$_2$ were prepared via a self-flux method, as described in Ref.~\cite{Ce5CoGe2single}. The heat capacity was measured using a Quantum Design Physical Property Measurement System (PPMS). Zero-field (ZF) $\mu$SR measurements were performed using the VMS spectrometer at the Swiss Muon Source (S$\mu$S) of the Paul Scherrer Institut (PSI), Switzerland.  Powdered single crystals were mounted on 25 $\mu$m thick Cu-foil, such that muons either were implanted in the sample or passed through. The sample was loaded in an Oxford Variox cryostat with a Heliox $^3$He insert with a base temperature of 0.3 K. Spin-polarized positive muons are implanted in the sample, and the decay positrons are preferentially emitted along the muon-spin direction, which are detected at either forward or backward positions. The asymmetry of the emitted positrons is calculated using 
\begin{equation}
	A(t)=\frac{N_F-\alpha N_B}{N_F+\alpha N_B},
\end{equation}
where the calibration constant $\alpha$ is determined from measurements in a weak-transverse field. All the $\mu$SR data were analyzed using the \musrfit\ software package \cite{musrfit}. Since there is an artifact in the data of one of the detectors at around 0.035 $\mu$s, the data in the range 0.025-0.04 $\mu$s has been excluded when fitting, and is not presented. Powder neutron diffraction measurements were performed on the WISH diffractometer at the ISIS facility of the Rutherford Appleton Laboratory, UK \cite{wish}. Analysis of the magnetic space groups (MSG)  was performed using \jana\ \cite{Jana2020}. The  \maxmagn\ \cite{MAXMAGN} software was used to generate magnetic model input files, and the refinements of the crystal and magnetic structures were performed using \fullprof\ \cite{Fullprof}. Crystalline-electric field (CEF) calculations based on a point charge model were performed with \pycrystalfield\ \cite{PyCrystalField}, and the Dzyaloshinskii–Moriya (DM) interaction analysis was performed using \spinw\ \cite{spinw}.
	
\section{Results\label{Results}}
\subsection{Specific heat and isothermal magnetization\label{HC}}

\begin{figure}[htbp]
\includegraphics[width=0.42\textwidth]{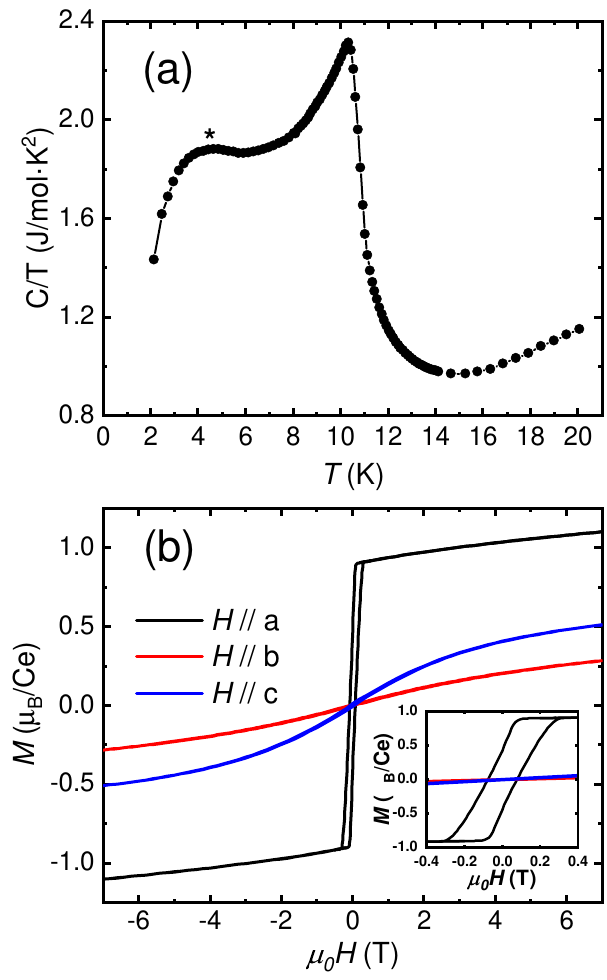}
\caption{\label{fig1} 
(a) The specific heat $C(T)/T$ of Ce$_5$CoGe$_2$, with a clear transition at $11$ K and a broad hump at $ \simeq 4.5 - 5$ K. (b) Isothermal magnetization $M(H)$ curves for $H \parallel a$, $H \parallel b$ and $H \parallel c$-axis, where a low-field loop appears only for magnetic fields along the $a$ axis. A magnified plot of the hysteresis loop is shown in the inset..
}
\end{figure}

The specific heat as $C/T$ of \ce\ is shown in Fig.~\ref{fig1}(a), which exhibits a sharp magnetic transition  at $T_{\textrm{C}}=11$ K, consistent with the previous reports \cite{Ce5CoGe2poly,Ce5CoGe2single}.  At lower temperatures, there is a broad hump centered at around 4.5 to 5 K (black asterisk), which could be related to the formation of short-range correlations or spin fluctuations. Figure~\ref{fig1}(b) displays the isothermal magnetization curves $M(H)$ at 2 K with a magnetic field applied along the crystallographic $a$, $b$ and $c$ axes. Clear hysteresis  between the up- and down-field sweeps is observed for $H \parallel a$ while no obvious hysteresis is present for $H \parallel b$ and $H \parallel c$, indicating that the Ce moments are predominantly orientated along the $a$-axis. For $H \parallel a$, $M(H)$ reaches a value of 1.1~$\mu_{\textrm{B}}$/Ce at $\mu_0H = 7$~T, consistent with previous $M(H)$ and $M(T)$ measurements \cite{Ce5CoGe2single}. 

\subsection{Muon spin relaxation\label{muSR}}
	
\begin{figure}[htbp]
\includegraphics[width=0.47\textwidth]{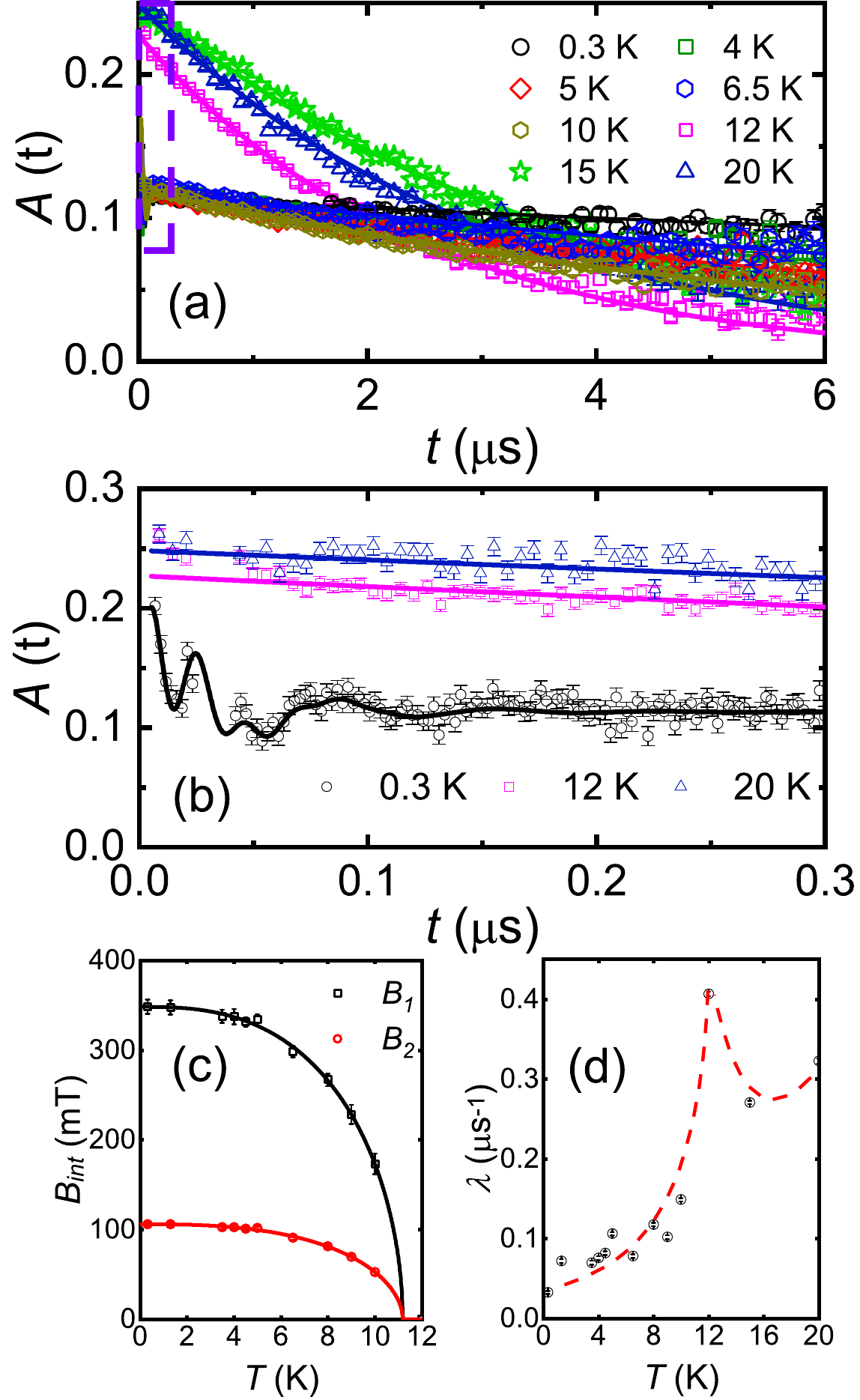}
\caption{\label{figmuSR}(a) Zero-field $\mu$SR spectra up to 6 $\mu$s  at several temperatures up to 20~K. (b) Data at low-times below 0.3 $\mu$s, corresponding to the dashed box in panel (a), measured at 0.3, 12 and 20 K. Clear multi-component high-frequency oscillations emerge below $T_{\textrm{C}}$, which can be resolved at short times. (c) Temperature evolution of the internal magnetic field components $B_1$ and $B_2$ obtained from fitting to Eq.~\ref{eqmag}. The solid lines show the fits using Eq.~(\ref{orderparameter}). (d) Temperature evolution of the exponential relaxation rate $\lambda$ (Eqs.~\ref{eqpm} and \ref{eqmag}). The red dashed line is a guide to eyes.}
\end{figure}
In order to understand the temperature evolution of the magnetic state as well as spin dynamics of \ce, ZF-$\mu$SR measurements were performed on powdered single crystals since ZF-$\mu$SR is very sensitive to magnetic order and spin dynamics \cite{RMPmusr}. Figure~\ref{figmuSR}(a) displays the $\mu$SR time spectra at several temperatures between 0.3 and 20 K. Above 11 K, \ce\ is in the paramagnetic state and the relaxation of the $\mu$SR asymmetry can be described by
\begin{equation}\label{eqpm}
A(t) = A_0e^{-\lambda t},
\end{equation}
\noindent where $\lambda$ is the Lorentzian relaxation rate and $A_0$ is the initial asymmetry corresponding to the exponential decay.  Below 10~K, there is a rapid relaxation of the asymmetry at low times, and high frequency oscillations appear below 0.3~$\mu$s.  As shown in Fig.~\ref{figmuSR}(b), the periodicity of the oscillations is not consistent with a single harmonic oscillation, indicating the presence of multiple components. Therefore, the spectra below $T_{\textrm{C}}$ were analyzed using two oscillatory components and an exponential relaxation :
\begin{equation}\label{eqmag}
A(t) = A_0e^{-\lambda t}+\sum_{i=1}^2 A_ie^{-\lambda_i t}\cos(\gamma_\mu B_i t+\phi),
\end{equation}
\noindent  where $A_i$ are the amplitudes of the oscillatory components, $B_i$ and $\lambda_i$ are the magnetic fields and the corresponding Lorentzian relaxation rates,  $\phi$ is a common phase, and $\gamma_{\mu}/2\pi = 135.5\, \mbox{MHz/T}$ is the muon gyromagnetic ratio \cite{PRBmusr}.  The total initial asymmetry $A_{\rm{tot}} = A_0+\sum A_i$ is fixed from fitting the data in the paramagnetic state using Eq.~\ref{eqpm} and the ratio $B_1/B_2$ = 3.25 is fixed based on a global fitting of all the data below $T_{\textrm{C}}$. The fitting using Eq.~\ref{eqmag} is shown by the solid lines in Fig.~\ref{figmuSR}(b), and the temperature evolution of the two internal field components $B_{\rm{int}}=B_1, B_2$ is displayed in Fig.~\ref{figmuSR}(c). Note that there is no anomaly in these internal fields at the temperature range of the broad hump in $C/T$ (Fig.~\ref{figmuSR}), suggesting this feature is not related to changes in the structure of the static order. This temperature dependence  is  fitted to the phenomenological function  \cite{musrblundell,magnetismblundell}
\begin{equation} \label{orderparameter}
	B_{\mbox{int}}(T) = B_{\mbox{int}}(0) [1-(T/T_{\textrm{C}})^\alpha]^\beta
\end{equation}
\noindent Here $B_1$ and $B_2$ were fitted with common values of $T_{\textrm{C}}$, $\alpha$ and $\beta$, as  shown in Fig.~\ref{figmuSR}(c), yielding $B_1(0)=348(1)$ mT,  $B_2(0)=106(6)$ mT, $\alpha=2.7(2)$, $\beta=0.52(6)$ and $T_{\textrm{C}} = 11.2(2)$ K. The fitted $T_{\textrm{C}}$ is consistent with specific heat measurements and the critical exponent $\beta$ is close to the value of 0.5 expected for a mean field magnet \cite{magnetismblundell}, while the large $\alpha$ values could imply  complex interactions between magnetic moments. Figure~\ref{figmuSR}(d) shows the temperature dependence of the Lorentzian relaxation rate $\lambda$, which exhibits 
a pronounced peak at the magnetic transition, consistent with a critical slowing down of spin fluctuations approaching $T_{\textrm{C}}$.

\subsection{Neutron diffraction\label{neutron}}
\begin{figure}[h]
	\includegraphics[width=0.48\textwidth]{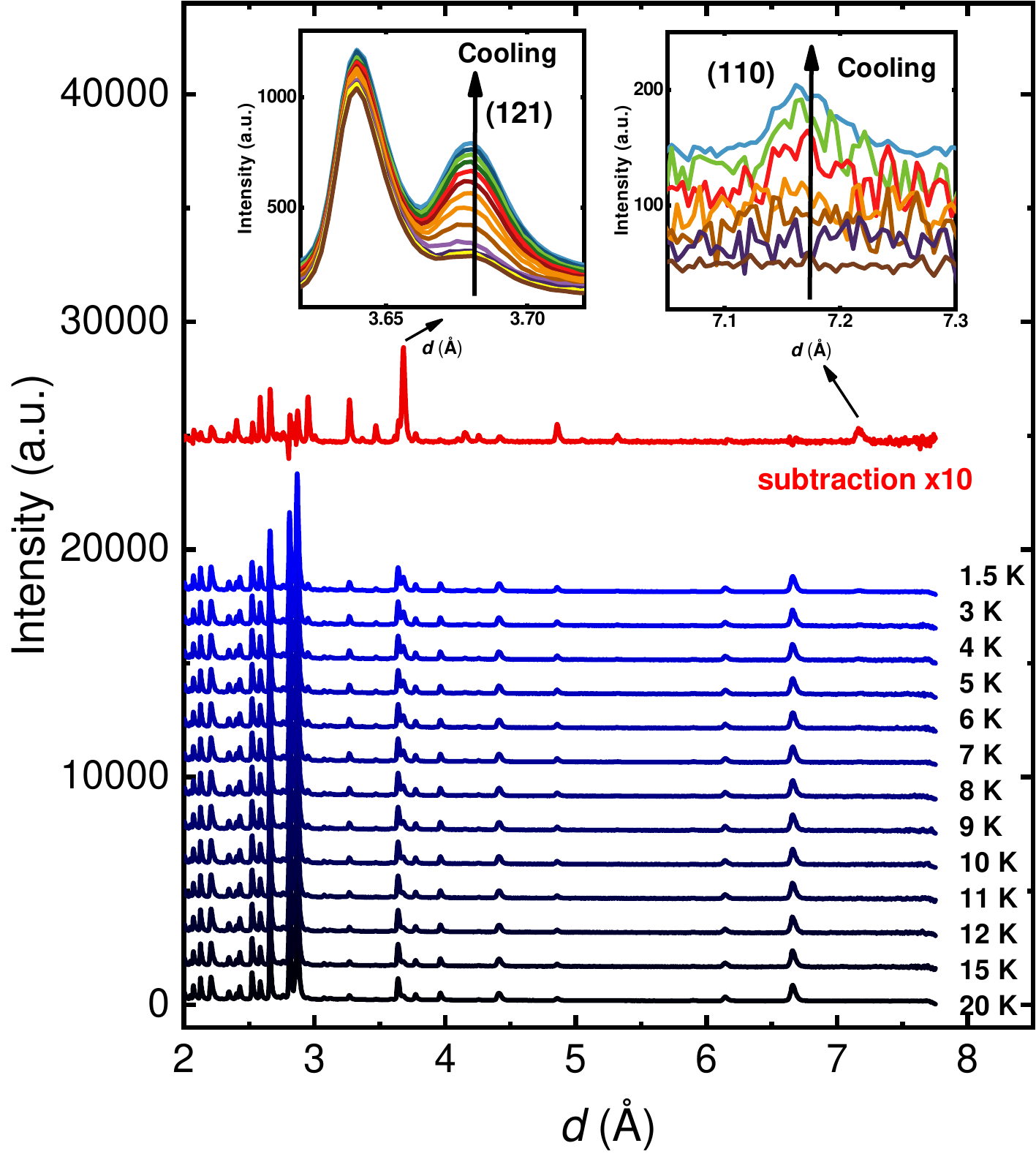}
	\caption{\label{neutrondata} Low-temperature neutron diffraction patterns of Ce$_5$CoGe$_2$ measured using the WISH diffractometer.  The red curve shows the difference plot from subtracting the 20~K data from that at 1.5~K, with the intensity enlarged by a factor of 10 so as to visualize the magnetic Bragg peaks. The insets show the temperature evolution of the (121) and (110) Bragg peaks.}
\end{figure}

To further understand the microscopic nature of the magnetic ground state, \ce\ has been investigated by neutron powder diffraction measurements. The neutron powder diffraction patterns at temperatures ranging from 1.5 to 20 K are shown in Fig.~\ref{neutrondata}. Upon cooling below $T_{\textrm{C}}$, there is additional intensity on some nuclear Bragg peaks. Since all the magnetic Bragg peaks can be indexed by integer ($HKL$), the propagation vector is determined to be $\mathbf{k}=0$, indicating that  the magnetic structure has the same periodicity as the crystallographic unit cell, as expected for the FM order observed in the magnetization (see Fig.~\ref{fig1}(b)). The pure magnetic contribution is  obtained by subtracting the diffraction pattern measured at 20 K from that at 1.5~K, which is shown by the red curve in Fig.~\ref{neutrondata}. The top insets show the temperature evolution of two integer ($HKL$) peaks. The (121) peak corresponds to a weak structural peak, and the intensity increases significantly below $T_{\textrm{C}}$, while the $(110)$ peak at long $d$-spacing is absent in the paramagnetic state, but has detectable intensity in the magnetically ordered phase.

Since the irreducible representations (irreps) are one dimensional real irreps, there is a one-to-one correspondence between the irreps of the parent space group $Pnma$ and the magnetic space groups (MSG) of the same geometry class \cite{MSG1,MSGJana2020}. Therefore, we performed the magnetic symmetry analysis using MSG  to describe and refine the magnetic structure. There are eight possible MSGs: $Pnma$, $Pn'm'a$, $Pnm'a'$, $Pn'ma'$, $Pn'm'a'$, $Pnma'$, $Pn'ma$ and  $Pnm'a$, and in the crystal  structure of \ce\  the Ce atoms occupy four inequivalent crystallographic sites, with Ce1, Ce2 and Ce3 occupying  sites with the $4c$ Wyckoff position, and Ce4 in a  $8d$ site. Note that since La$_5$CoGe$_2$ is nonmagnetic,  only magnetic moments on Ce atoms are considered in the analysis \cite{Ce5CoGe2single,Ce5CoGe2poly}.

For all possible MSGs, three components of the magnetic moment $(m_x,m_y,m_z)$ are allowed by symmetry for Ce4, while for Ce1, Ce2 and Ce3, four MSGs only have $y$ components allowed (described by $(0,m_y,0)$), while for the other MSGs the $y$ components are forbidden (described by $(m_x,0,m_z)$).
Moreover only the MSGs $Pn'm'a$, $Pnm'a'$ and $Pn'ma'$ allow  overall net magnetic moments, and only $Pnm'a'$ allows a net magnetic moment along the $a$-axis.

\begin{figure}[h]
	\includegraphics[width=0.45\textwidth]{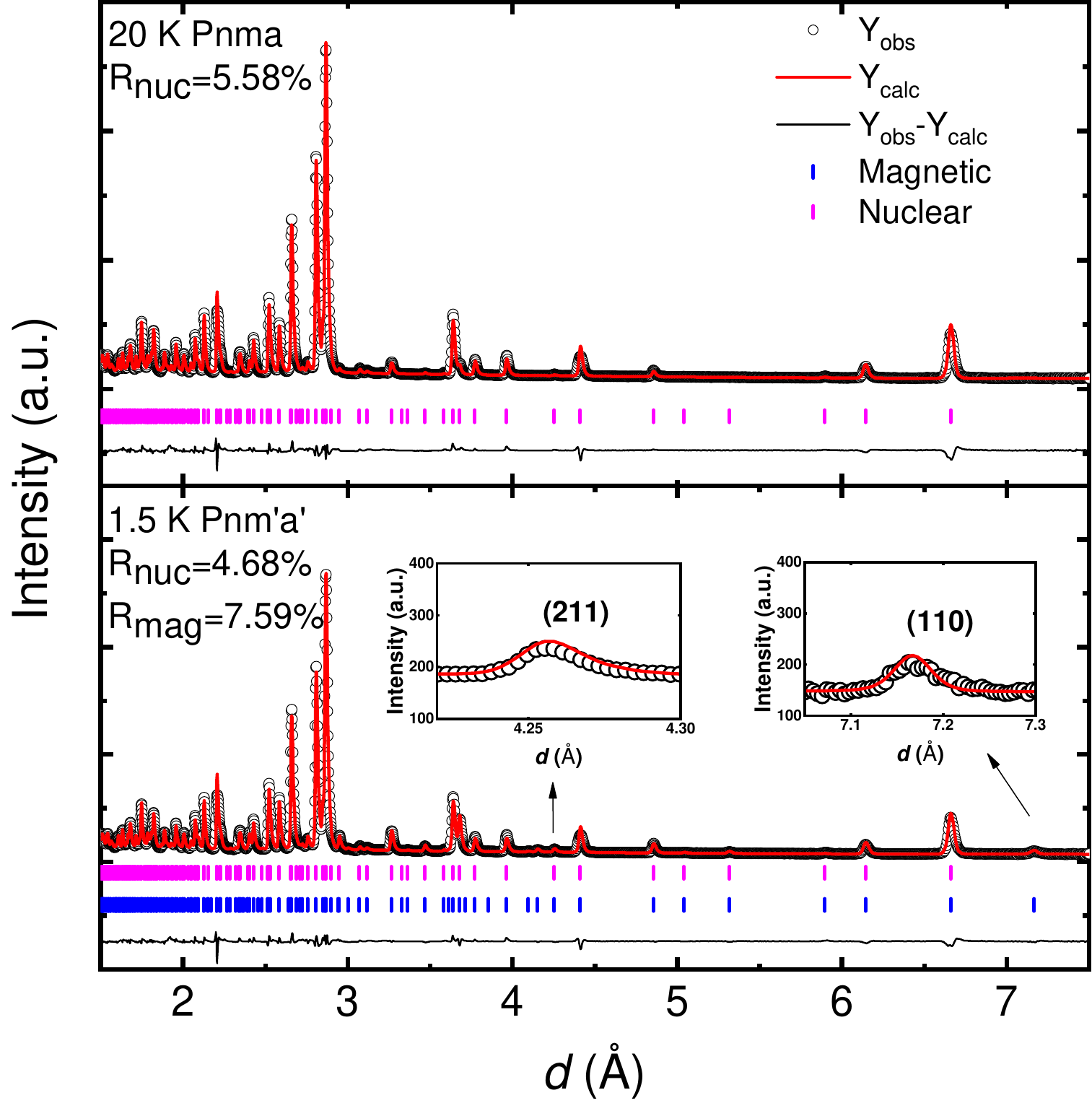}
	\caption{\label{neutronrefinement} Neutron diffraction patterns in the paramagnetic phase (20 K) and at the magnetic phase (1.5 K), along with the calculations based on refinements of the crystal structure, as well as the magnetic structure with the  $Pnm'a'$ MSG. The two insets in the bottom panel magnify the (121) and (110) peaks,  which correspond to the black asterisks in the bottom panel. The experimental, calculated, and difference patterns are shown as black circles, red curves, and black curves, respectively. The magnetic and nuclear Bragg peaks are indicated by vertical blue and pink bars, respectively.}
\end{figure}

As shown in the top panel of Fig.~\ref{neutronrefinement}, the pattern at 20 K is well refined incorporating only the crystal structure with the  $Pnma$ space group. Refinement to the eight possible MSGs were then performed to determine the magnetic structure of \ce (see Appendix~\ref{Appendix2}). Only the magnetic structure corresponding to $Pnm'a'$ could well describe the observed magnetic peaks, with the refinement shown in the bottom panel of Fig.~\ref{neutronrefinement}, which is consistent with the FM behavior along the $a$~axis found in $M(H)$ measurements in Fig.~\ref{fig1}. The lattice parameters from refinements of the crystal structure at 1.5 K are $a=12.2840(3)$\AA, $b=8.81795(18)$\AA, $c=7.92290(17)$\AA, consistent with Ref. \cite{Ce5CoGe2single}. The parameters from the refinements  of the crystal and magnetic structures are displayed in Table.~\ref{magstr}. Importantly, although the refined magnetic structure has a net moment along the $a$~axis, it does not correspond to a simple collinear ferromagnetic structure, and the data cannot be refined fixing the symmetry-allowed $b$ or $c$ components of the moments to zero. In particular, a nonzero component along the $b$-axis  on Ce4 ($m_y$) is required to well describe the intensity of the (211) peak, while only components along the $c$ axis ($m_z$) contribute to the intensity on the (110) peak (inset of Fig.~\ref{neutrondata}).  As shown in the insets of the bottom panel of Fig.~\ref{neutronrefinement}, refinements incorporating all the symmetry-allowed $b$ and $c$ components can well describe the  (211) and (110) peaks. Therefore all the symmetry allowed components are necessary in the refinement, necessitating nine fitting parameters for the magnetic structure, but the presence of the aforementioned magnetic peaks makes the conclusions of noncollinear ferromagnetism robust.  Note also that the data cannot be accounted for with the same moment on each of the Ce-sites. The refinements also reveal a significant variation of the ordered moments between different Ce-sites, where the average  net moment value is 0.98(2)\,$\mu_{\textrm{B}}$/Ce, consistent with 1.1\,$\mu_{\textrm{B}}$/Ce from $M(H)$ measurements.

\begin{figure}[htp]
\includegraphics[width=0.45\textwidth]{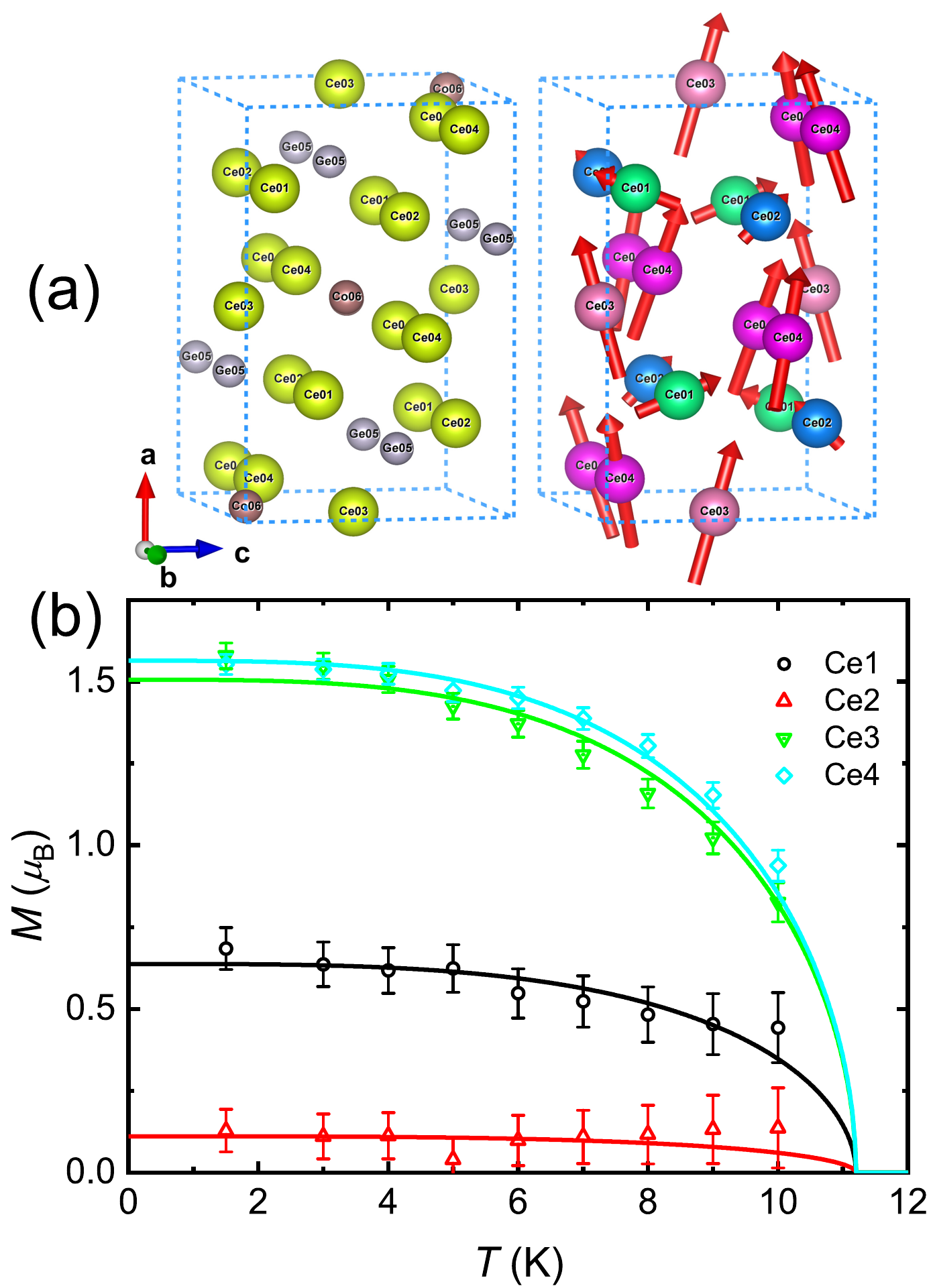}
\caption{\label{neutronmoment} (a) Refined crystal  (left) and magnetic  (right) structures of Ce$_5$CoGe$_2$. Only Ce atoms are shown in the right panel. (b) Temperature dependence of the moments of the four Ce sites obtained from refinements of neutron diffraction data, where solid lines show fits to Eq.~\ref{orderparameter}.}
\end{figure}

The refined crystal and magnetic structures are displayed in Fig.~\ref{neutronmoment}(a), where the different sites are labelled. It can be seen that Ce3 and Ce4 form a layer perpendicular to the $a$-axis with larger predominantly $a$-axis moments ($\sim1.5\mu_{\textrm{B}}$/Ce), while the Ce1 and Ce2 layers have smaller moments, with comparable values of $m_x$ and $m_z$. The temperature dependences of the total moments per Ce for the four sites are  displayed in Fig.~\ref{neutronmoment}(b), where they all show similar behavior to $B_1$ and $B_2$ in Fig.~\ref{figmuSR}, saturating below about $4\,\mbox{K}$. No obvious change in magnetic structure or moment components are found  around 4.5 to 5 K. The solid lines display the fitting to $M(T) = M(0) [1-(T/T_{\textrm{C}}^\alpha]^\beta$ \cite{musrblundell,magnetismblundell} with $T_{\textrm{C}}$ and $\beta$ fixed from the fitting of $B_1$ and $B_2$.

\section{Discussion and summary\label{Dis}}

Our neutron diffraction measurements reveal an unusual noncollinear FM structure in \ce, the origin of which is to be determined. Although the $M(H)$ shows clear ferromagnetic loops for fields along the $a$~axis [Fig.~\ref{fig1}(b)], the magnetization is not fully saturated in a 6~T field, which may reflect the canting of the noncollinear moments towards the $a$~axis in applied magnetic fields. Determining the influence of CEF is particularly important for understanding the magnetic ground state, especially given the presence of multiple inequivalent sites. Within the crystal structure of \ce\ with orthorhombic point group $Pnma$, the Ce ions have two local point group symmetries: $C_s$ (mirror plane, monoclinic) for the $4c$ sites (Ce1, Ce2 and Ce3) and  $C_1$ (identity operator, triclinic) for the $8d$ site (Ce4). Note that to simplify the calculations, we use the \pycrystalfield\ formalism which prioritizes having the $y$ axis normal to a mirror plane  \cite{PyCrystalField}, rather than the formalism in Walter \cite{CEFWalter} which prioritizes the $z$ axis being along a rotation axis. In this case, the CEF Hamiltonian for Ce1, Ce2 and Ce3 has the form:
\begin{align}\label{4c}
	H_{\textrm{CEF}}^{4c} = &B_2^0 O_2^0 + B_2^1 O_2^1 + B_2^2 O_2^2 + B_4^0 O_4^0\notag\\
	\quad + &B_4^1 O_4^1 + B_4^2 O_4^2 + B_4^3 O_4^3 + B_4^4 O_4^4 
\end{align}
\noindent where $B_n^m$ are Stevens parameters and $O_n^m$ are the Stevens operator equivalents \cite{CEFstevens}. From diagonalizing the CEF Hamiltonian to obtain the eigenvectors \cite{CEFstevens}, the expectation value of the $y$ axis moment $\langle \mu_y \rangle = g_J\psi^{\dagger}\hat{J}_y\psi = 0$ is calculated, which is a result of the aforementioned mirror symmetry, and is in line with the refined magnetic structure with $m_y=0$.

For Ce4 with only $C_1$ symmetry in the \pycrystalfield\ formalism \cite{PyCrystalField}, the CEF Hamiltonian is 
\begin{align}\label{8d}
	H_{\textrm{CEF}}^{8d} = &B_2^0 O_2^0 + B_2^{\pm 1} O_2^{\pm 1} + B_2^{\pm 2} O_2^{\pm 2}+ B_4^0 O_4^0 \notag\\
	\quad  + &B_4^{\pm 1} O_4^{\pm 1} + B_4^{\pm 2} O_4^{\pm 2} + B_4^{\pm 3} O_4^{\pm 3} + B_4^{\pm 4} O_4^{\pm 4} 
\end{align}

\begin{table*}[htbp]
	\centering
	\caption{\label{magstr}Refined nuclear and magnetic structure parameters of \ce\  based on MSG $Pnm'a'$ at 1.5 K. The expectation values of corresponding moments based on the point charge model are also shown.}
	\renewcommand\arraystretch{1.3}
	\begin{tabularx}{\textwidth}{sssssssss} 
		\toprule[1pt]
		\toprule
		\multicolumn{9}{c}{\textbf{Nuclear structure (Space group $Pnma$)}}  \\
		\cmidrule{1-9}
		Atom & Wyck. & $x$ & $y$ & $z$ & Latt. Para. & Value \\
		\midrule
		Ce1 & 4c & 0.7150(8) & 0.25000 & 0.6673(10) & $a($\AA$)$ & 12.2840(3) \\
		Ce2 & 4c & 0.2889(7) & 0.25000 & 0.3469(9) & $b($\AA$)$ & 8.81795(18) \\
		Ce3 & 4c & 0.4973(7) & 0.25000 & 0.9652(7) & $c($\AA$)$ & 7.92290(17) \\
		Ce4 & 8d & 0.4337(3) & 0.5296(5) & 0.6811(5) & $\alpha(^{\circ})$ & $90$ \\
		Ge  & 4c & 0.67095(15) & 0.5027(5) & 0.9358(3) & $\beta(^{\circ})$ & $90$ \\
		Co  & 8d & 0.4797(9) & 0.25000 & 0.5637(13) & $\gamma(^{\circ})$ & $90$ \\
		\midrule
		\multicolumn{5}{c}{\textbf{Magnetic structure (MSG $Pnm'a'$)}} & \multicolumn{4}{c}{\textbf{Point charge model expectation}} \\
		\cmidrule(r){1-5} \cmidrule(l){6-9}
		Atom & $m_x$ & $m_y$ & $m_z$ & Net moment per unit cell & $\langle \mu_x \rangle$ & $\langle \mu_y \rangle$ & $\langle \mu_z \rangle$ & Net moment per unit cell \\
		\midrule
		Ce1 & 0.25(6) & 0.00 & 0.62(6) & (1.00,0,0) & $0.26$ & 0.00 & $0.64$ & (1.04,0,0) \\
		Ce2 & 0.18(6) & 0.00 & 0.18(6) & (0.72,0,0) & $0.12$ & 0.00 & $1.37$ & (0.48,0,0) \\
		Ce3 & 1.52(4) & 0.00 & -0.43(5) & (6.08,0,0) & $0.97$ & 0.00 & $0.82$ & (3.88,0,0) \\
		Ce4 & 1.48(3) & 0.52(4) & 0.39(3) & (11.84,0,0) & $1.36$ & $0.55$ & $0.58$ & (10.88,0,0) \\
		\bottomrule
		\bottomrule
	\end{tabularx}
\end{table*}

The presence of four inequivalent Ce sites, all with low symmetry, makes experimental determination of the CEF parameters challenging, and therefore to gain an understanding a point charge model was used to estimate the CEF parameters (see Appendix~\ref{Appendix}). Table.~\ref{magstr} displays the expectation values of the three components of the magnetization based on the point charge model where +1.6, -2, -3 are used as the valence states for Ce, Co and Ge, respectively. Note that these values of valence are empirical parameters chosen such that  the observed variations  of the magnetic moments between different Ce sites are qualitatively reproduced, while maintaining charge balance. It can be seen that the expectation values of the magnetic moments for Ce1 and Ce4 are quite close to the experimental results, whereas for Ce2 and Ce3 $\langle \mu_z \rangle$ is larger than the refined values of $m_z$. While these discrepancies could be a result of the simplified nature of the point charge model, it may also reflect competition between the single-ion anisotropy of these sites favoring the moment pointing away from the $a$-axis, and FM exchange with the Ce1 and Ce4 sites favoring alignment along the $a$-axis, and this frustration between magnetic exchange and CEF effects is a possible origin of the non-collinear ferromagnetism. Moreover, while the effective moment determined from the high temperature magnetic susceptibility is 2.39\,$\mu_{\textrm{B}}$/Ce \cite{Ce5CoGe2single}, which is close to the free ion value, the estimated Kondo temperature $T_{\textrm{K}}$ = 14.2 K is comparable to the magnetic transition temperature $T_{\textrm{C}}$ = 10.9 K, and therefore Kondo screening of the local moments by the conduction electrons may also reduce the magnetic moments below the expected values for the CEF ground states. However, a point charge model which assumes the electron density is situated only on the surrounding atoms is often unable to numerically reproduce the CEF level schemes of correlated intermetallic compounds \cite{Buschow1980}, and therefore a precise comparison between the sizes of the ordered moments and those corresponding to the ground state CEF wave function require experimental determination of the CEF parameters. Nevertheless, these calculations suggest that differences in the local environments and CEF potentials between the different inequivalent Ce sites can qualitatively explain there being significant differences in their ordered moments. Note also that in these calculations, $m_y=0$ for the $4c$ sites, in line with the above symmetry arguments. This also constrains the magnetic structure to belong to one of the 4 MSGs where non-zero $m_x$ and $m_z$ are allowed by symmetry.

Another mechanism for noncollinear magnetic structures is the DM interaction. Although \ce\ crystallizes in a centrosymmetric crystal structure, whether DM interactions are allowed depends on the point group symmetry of the bond center.  We have analyzed whether DM interactions are symmetry allowed for  bonds between Ce ions using \spinw\,\cite{spinw}.  While the DM vector for the nearest neighbor bond between two Ce4 atoms is enforced to be zero,  of the 19 types of bonds with separations up to 5\,\AA, only two bonds lie at an inversion center, and therefore most bonds can potentially have a nonzero DM vector. As a result, the DM interaction could also play a role in enabling a non-collinear magnetic structure, as is the case in compounds hosting magnetic skyrmion and meron lattices \cite{Skyrmion1,Skyrmion2,Meron2,MeronCeAlGe}.

In summary, we performed neutron diffraction and $\mu$SR measurements to gain a microscopic understanding of the magnetic structure and spin dynamics of \ce. Neutron diffraction reveals a  noncollinear ferromagnetic structure, with different moments on the four inequivalent Ce sites, which may arise due to frustration between the single ion anisotropies arising from the CEF, intersite RKKY and DM magnetic interactions, and Kondo screening. Given that there is a change of magnetic ground state to probable AFM order in \ce\ under pressure \cite{zhang2026pressureinducedsuperconductivitymagneticquantum}, these results suggest that this change could be driven by the pressure evolution of these frustrated components that already influence the ambient pressure magnetic properties, indicating that probing the magnetic structures of other correlated ferromagnetics exhibiting such a scenario under pressure would be warranted. \cite{Sidorov2003,CeRuPO,Friedemann2018,AvoidedFerromagneticQuantumlengyel2015}. Moreover, while other Ce$_5M$Ge$_2$ compounds 
such as those with $M=$Rh, Ru and Pd also exhibit magnetic transitions in a similar temperature range, with ferromagnetic-like hysteresis in the low temperature magnetization, the magnetization of the $M=$Pd and Ru compounds exhibit metamagnetic transitions \cite{Ce5PdGe2, Ce5RuGe2}, which are not observed in the Co and Rh compounds \cite{Ce5CoGe2poly, Ce5CoGe2single,Ce5RhGe2}. The lack of a metamagnetic transition in \ce\ is consistent with our neutron diffraction results, since the net magnetic moments of all four Ce sites in the deduced magnetic structure are coaligned, and therefore a field-induced spin-flip transition is not anticipated. On the other hand, Ce$_5$PdGe$_2$ and Ce$_5$RuGe$_2$ may exhibit ferrimagnetic or other partially spin-compensated magnetic structures, possibly due to different orientations of the sublattice magnetization for different Ce-sites, but this needs to be confirmed by neutron diffraction.
Since the magnetic ground state in \ce\ does not correspond to simple uniform ferromagnetism, these findings highlight the importance of microscopic characterizations of the ordered  states of correlated ferromagnets, which can be related to the outcome upon tuning the material to quantum phase transitions. To this end, it is of particular interest to examine the effects of tuning the competing interactions in \ce\ using parameters such as doping, element substitution, or strain.

\section{Acknowledgments}
We are grateful to Pascal Manuel for assistance with performing the neutron diffraction experiment. Work at Zhejiang University was supported by the National Key R\&D Program of China (Grant No. 2022YFA1402200, 2023YFA1406303), the National Natural Science Foundation of China (Grant No. W2511006, 12034017, 12494592, U23A20580). The neutron diffraction is based on experiment No. RB2420292 \cite{Ce512WISH} performed at the ISIS neutron and muon source, UK and the muon experiments are based on experiment No. 20231320 at the VMS spectrometer at Swiss Muon Source S$\mu$S, Paul Scherrer Institute, Villigen, Switzerland.
	
\appendix

\section{CEF parameters for the point charge model\label{Appendix}}
The Stevens parameters $B_n^m$  calculated using a point charge model with  +1.6, -2, -3  as the valence states for Ce, Co and Ge, respectively, are shown in Table.~\ref{Bnm}. The wavefunctions of the corresponding ground state doublets are also displayed in Table.~\ref{GS}.
\begin{table*}
	\caption{Calculated values of the Stevens parameters $B_n^m$ for Ce$_5$CoGe$_2$ derived from the  point charge model with +1.6, -2, -3 used as the valence states for Ce, Co and Ge.\label{Bnm}}
	\begin{ruledtabular}
		\begin{tabular}{ccccccccccccccc}
			 & $B_2^{-2}$ & $B_2^{-1}$ & $B_2^{0}$ & $B_2^{1}$ & $B_2^{2}$ & $B_4^{-4}$ & $B_4^{-3}$ & $B_4^{-2}$ & $B_4^{-1}$ & $B_4^{0}$ & $B_4^{1}$ & $B_4^{2}$ & $B_4^{3}$ & $B_4^{4}$ \\
			\midrule
			Ce1 & 0.000 & 0.000 & -0.833 & 20.313 & 10.015 & 0.000 & 0.000 & 0.000 & 0.000 & -0.079 & -0.111 & -0.413 & 2.494 & 0.159 \\
			Ce2 & 0.000 & 0.000 & -4.001 & 16.492 & 14.773 & 0.000 & 0.000 & 0.000 & 0.000 & -0.093 & 0.083 & -0.231 & 2.313 & 0.052 \\
			Ce3 & 0.000 & 0.000 & 9.962 & 27.187 & 1.615 & 0.000 & 0.000 & 0.000 & 0.000 & 0.112 & 0.386 & 0.089 & 1.209 & -0.963 \\
			Ce4 & -1.037 & 5.618 & -1.053 & -1.505 & -6.893 & 0.167 & -0.641 & 0.170 & -0.349 & -0.002 & -0.303 & 0.072 & 1.652 & 0.358 \\
		\end{tabular}
		
	\end{ruledtabular}
	\label{tab:Bnm}
\end{table*}

\begin{table*}
	\caption{Wavefunctions of the ground state doublets from the point charge model described in the text.\label{GS}}
	\begin{ruledtabular}
		\begin{tabular}{ccccccc}
			 & $| -\frac{5}{2}\rangle$ & $| -\frac{3}{2}\rangle$ & $| -\frac{1}{2}\rangle$ & $| \frac{1}{2}\rangle$ & $| \frac{3}{2}\rangle$ & $| \frac{5}{2}\rangle$ \\
			\midrule
			Ce1(1) & -0.270 & -0.702 & -0.226 & 0.612 & -0.088 & 0.000 \\
			Ce1(2) & 0.000 & -0.088 & -0.612 & -0.226 & 0.702 & -0.270 \\
			Ce2(1) & 0.788 & 0.224 & -0.482 & -0.055 & 0.305 & 0.026 \\
			Ce2(2) & -0.026 & 0.305 & 0.055 & -0.482 & -0.224 & 0.788 \\
			Ce3(1) & -0.201 & -0.668 & -0.679 & 0.189 & -0.128 & -0.012 \\
			Ce3(2) & -0.012 & 0.128 & 0.189 & 0.679 & -0.668 & 0.201 \\
			Ce4(1) & -0.068 & 0.237-0.200i & 0.079+0.555i & -0.065-0.373i & -0.151+0.394i & -0.303-0.414i \\
			Ce4(2) & 0.513 & 0.229-0.355i & 0.340-0.168i & 0.495-0.264i & 0.021-0.310i & -0.040-0.055i \\
		\end{tabular}
		
	\end{ruledtabular}
	\label{flo:Eigenvectors}
\end{table*}

\section{Fitting to the subtraction data\label{Appendix2}}

\begin{figure*}[!htb]
	\includegraphics[width=\textwidth]{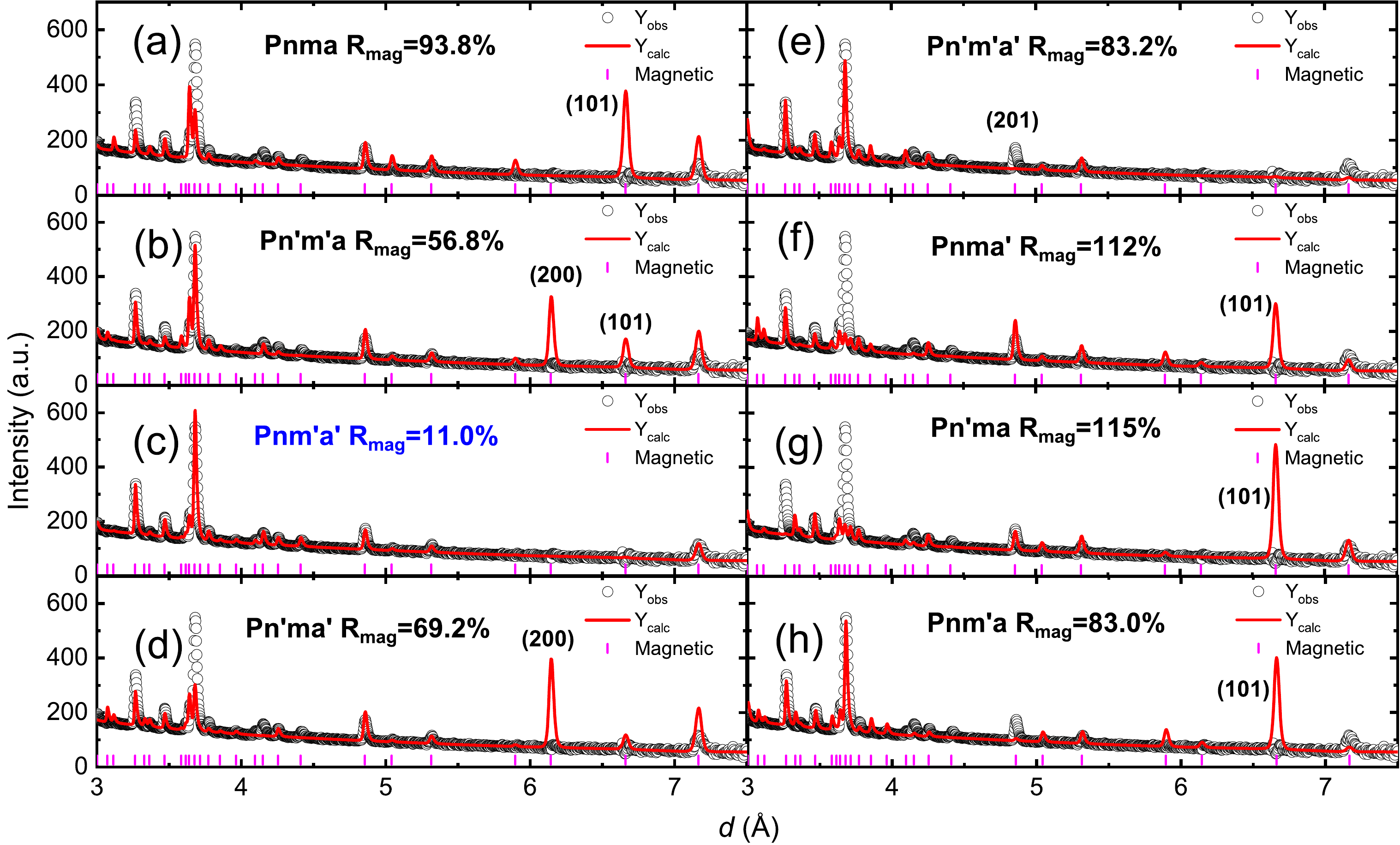}
	\caption{\label{subtraction fit} (a)-(e) Results of Rietveld refinements to the magnetic contribution of the diffraction data considering magnetic structures corresponding to the  magnetic space groups $Pnma$, $Pn'm'a$, $Pnm'a'$, $Pn'ma'$, $Pn'm'a'$, $Pnma'$, $Pn'ma$ and $Pnm'a$, respectively. The data is obtained by subtracting the 20 K data from that at 1.5 K and is shifted vertically by a constant value.  Note that the $R_{\textrm{mag}}$ for $Pnm'a'$ is a bit higher than in Fig.~\ref{neutronrefinement}, since for the latter the magnetic and nuclear contributions are simultaneously refined.}
\end{figure*}

To estimate the magnetic contribution to the scattering, the diffraction data collected in the paramagnetic state (at 20 K) were subtracted from that measured at 1.5 K. Subsequent Rietveld refinements were performed considering magnetic structures corresponding to the possible magnetic space groups, including $Pnma$, $Pn'm'a$, $Pnm'a'$, $Pn'ma'$, $Pn'm'a'$, $Pnma'$, $Pn'ma$ and $Pnm'a$. As displayed in Fig.~\ref{subtraction fit}, the refinements show that several groups, specifically  $Pnma$, $Pn'm'a$, $Pnma'$, $Pn'ma$ and $Pnm'a$,  predict a sizeable magnetic Bragg peak at the (101) position, which is absent in the diffraction pattern. Meanwhile, the groups $Pn'm'a$ and $Pn'm'a'$ incorrectly assign significant intensity to the (200) peak, while $Pn'm'a'$ fails to accurately describe the (201) peak. Only the magnetic space group $Pnm'a'$ (Fig.~\ref{subtraction fit}(c)) yields a satisfactory fit to the observed pattern, successfully reproducing all the key features with a significantly smaller magnetic R factor $R_{\rm mag}$, resulting in a magnetic structure with net moments aligned along the $a$-axis.

\end{document}